\documentclass[%
reprint,
superscriptaddress,
nofootinbib,
 amsmath,amssymb,
 aps,
 prc,
]{revtex4-1}

\usepackage{graphicx}
\usepackage{dcolumn} 
\usepackage{bm}      
\usepackage{xcolor}
\usepackage{dcolumn}
\newcolumntype{d}{D{.}{.}{-1}}
\usepackage{multirow}
\usepackage[colorlinks,citecolor=blue]{hyperref}
\newcommand{\fm}{\mathrm{\, fm}}
\newcommand{\I}{\mathrm{i}}
\newcommand{\MeV}{\mathrm{\, MeV}}
\renewcommand{\vec}[1]{\bm{#1}}

\begin{document}

\title{A Survey of Nuclear Pasta in the Intermediate Density Regime:\\
Structure Functions for Neutrino Scattering}

\author{B. Schuetrumpf}
\affiliation{GSI Helmholzzentrum f\"ur Schwerionenforschung, Planckstra\ss{}e 1, 64291 Darmstadt, Germany}
\author{G. Mart\'{i}nez-Pinedo}
\affiliation{GSI Helmholzzentrum f\"ur Schwerionenforschung, Planckstra\ss{}e 1, 64291 Darmstadt, Germany}
\affiliation{Institut f\"ur Kerphysik, Technische Universit\"at Darmstadt, Schlossgartenstra\ss{}e 2, 64289 Darmstadt, Germany}
\author{P.-G. Reinhard}
\affiliation{{Institut f{\"u}r Theoretische Physik II, Universit\"at Erlangen-N\"urnberg, 91058 Erlangen, Germany}}

\date{\today}

\begin{abstract}

\begin{description}
\item[Background] Nuclear pasta matter, emerging due to the
  competition between the long-range Coulomb force and the short-range
  strong force, is believed to be present in astrophysical scenarios,
  such as neutron stars and core-collapse supernovae. Its structure
  can have a high impact e.g. on neutrino transport or the tidal
  deformability of neutron stars.

\item[Purpose] We investigate the impact of nuclear pasta on neutrino
  interactions and compare the results to uniform matter.

\item[Method] We calculate the elastic and inelastic static structure
  factors for nuclear pasta matter using density functional theory
  (DFT), which contain the main nuclear input for neutrino
  scattering.

\item[Results] Each pasta structure leaves a unique imprint in the
  elastic structure factor and it is largely enhanced. The inelastic
  structure factors are very similar for all configurations.

\item[Conclusion] Nuclear pasta has a noticeable impact on neutrino
  neutral-current scattering opacities. While for inelastic reactions
  the cross section is reduced, the elastic coherent scattering
  increases dramatically. The effect can be of importance for the
  cooling of neutron stars as well as for core-collapse supernova
  models.

\end{description}
\end{abstract}
\maketitle
\section{Introduction}
Bulk matter with densities of about nuclear densities
($\rho_0=0.16\fm^{-3}$) is realized only in astrophysical objects,
such as neutron stars or in supernova explosions. These sites are
valuable laboratories, because nuclear matter properties, such as the
nuclear equation of state, greatly impact such astrophysical
scenarios. 

Nuclear matter around saturation density is well described as uniform
matter. However, at sub-saturation densities nuclear matter clusters
into a variety of intriguing shapes, commonly referred to as nuclear
pasta.  Nuclear pasta has been the topic of research for some
decades~\cite{Ravenhall,Hashimoto} and has been much studied with
classical~\cite{Dorso2012,Caplan2018} and quantum
methods~\cite{Williams1985,Oka13a,Pais15,Bonche,Mag02,Goe07a,NewtonStone,Pais12,Schuetrumpf2013,Schuetrumpf2014,Schuetrumpf2015,Schuetrumpf2015a,Fattoyev2017}. The
impact of pasta matter on astrophysical objects are manifold: the
viscosity of pasta matter can dampen the oscillations of neutron
stars, its thermal conductivity is important for the neutron star
cooling~\cite{Horowitz2008}, it can explain the limitation of spin
periods of pulsars~\cite{Pons2013} and its neutrino opacity has an
impact on the early stage of cooling of the neutron star and the
explodability of supernovae
\cite{Horowitz2004,Horowitz20042,Sonoda2007,Gry10,Roggero.Margueron.ea:2018,Nandi2018}. In
this paper, we focus on neutrino scattering and the impact of pasta
structures on neutrino opacities.

For neutrino scattering on uniform matter a large variety of studies
can be found in the literature (\emph{e.g.}
\cite{Reddy1998,Horowitz2002,Cowell2004,Burrows.Reddy.Thompson:2006,Pastore2015,Dzhioev2018,Shen2018,Bedaque2018,Riz2018}). They
are mostly based on density functional theory or effective
interactions and include in-medium particle correlations. These can
have a significant impact on the opacities and thus on the
astrophysical scenarios. For explicit nuclear pasta configurations the
neutrino opacity has only been studied using the molecular dynamics
(MD)
framework~\cite{Horowitz2004,Horowitz20042,Sonoda2007,Nandi2018}. MD
has the advantage that it includes real particle-particle correlations
between nucleons. However, the Pauli principle is only approximately
realized by a phenomenological potential. These studies show an
enhancement of neutrino scattering opacities due to coherent neutrino
scattering.

In this work, we will employ nuclear density functional theory to
calculate the structure factors of nuclear pasta, the main ingredients
for the neutrino cross section. We will focus on the pasta
configurations which appear at intermediate densities,
($\rho_0/4-\rho_0/2$) which have been discussed in an earlier
work~\cite{Schuetrumpf2019}. In section \ref{sec:method}, we introduce
the pasta configurations and explain briefly the computation of
neutrino scattering opacities and structure factors. In
\ref{sec:results}, we present the results for the various nuclear
pasta configurations and compare them to uniform matter.

\section{Nuclear Pasta and Neutrino Scattering}\label{sec:method}
\subsection{Minimal Surfaces Pasta Configurations}

\begin{figure}[htb]
    \centering
    \includegraphics[width=\columnwidth]{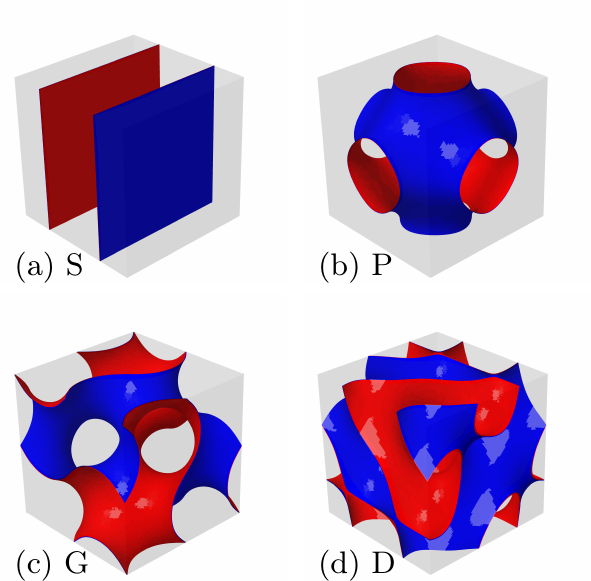}
    \caption{One unit cell of minimal surfaces indicating the dividing
      surface between matter and vacuum: (a) Slab, (b) P-surface, (c)
      G-surface, (d) D-surface}
    \label{fig:surfaces}
\end{figure}

In this work, we consider pasta configurations in the mid-density
range, between $0.04\fm^{-3}$ and $0.08\fm^{-3}$. All configurations
are related to minimal surfaces, which are the Primitive (P), the
Gyroid (G), and Diamond (D) surfaces
\cite{Michielsen2001,Schuetrumpf2013,Schuetrumpf2015}.  Furthermore we
also consider the flat or slab surface, which we label with (S). All
these surfaces minimize the surface area locally and have a vanishing
mean curvature. They can be characterized by the nodal approximations
\begin{subequations}
\label{eq:nodal_approx}
\begin{align}
     \phi_S &= \cos X  
     \;,
\\
     \phi_P &= \cos X + \cos Y + \cos Z 
     \;,
\\
     \phi_G &= \cos X \sin Y + \cos Y \sin Z + \cos Z \sin X
     \;,
\\
     \phi_D &= \cos X \cos Y \cos Z + \cos X \sin Y \sin Z \nonumber
     \;,
\\
            &\quad + \sin X \cos Y \cos Z + \sin X \sin Y \cos Z\;,
\end{align}
\end{subequations}
where $X=2\pi x/L$ and likewise for the other directions. $L$ is the
unit cell size. The condition $\phi_i=0$ defines the surfaces which
divide the space into two half spaces.  They are illustrated in
Fig.~\ref{fig:surfaces}. One side is colored in red (light), the other
in blue (dark). The division produces exactly two connected pieces. If
one (e.g. $\phi_i>0$) is filled with nuclear matter and the other
remaining (almost) void, then this is called a single configuration
because the space filled with matter is singly connected. For the case
of the Gyroid, we also consider double configurations. Here either
both half spaces are filled partly with nuclear matter ($|\phi_i|>t$)
with a volume around the Gyroid surface void. $t>0$ is a filling
parameter. We call this case double Gyroid network-like (dGn)
configuration. The inverted case, where nuclear matter is located in
the vicinity of the surface ($|\phi_i|<t$) is called double Gyroid
surface-like configuration (dGs).

Simple models for the densities can be generated using the idealized
dividing surfaces Eq.~(\ref{eq:nodal_approx}) with 
\begin{subequations}
\begin{align}
  \rho^\mathrm{(soft)}(\bm{r})&=\rho_0^\mathrm{(soft)}[\phi_i(\bm{r})-\phi_{i,\text{min}}]\label{eq:soft}\\ 
  \rho^\mathrm{(hard)}(\bm{r})&=\rho_0^\mathrm{(hard)}[\theta(\phi_i(\bm{r}))-\phi_0]\label{eq:hard}
\end{align}
\end{subequations}
with $i\in \{S,P,G,D,dGn,dGs\}$, $\theta(x)$ is the Heaviside theta
function and $\phi_{i,\text{min}}$ the minimum value of $\phi_i(\bm{r})$. We adjust $\rho_0$ such that the desired mean density is
reached. For the double surfaces we take the functions
\begin{subequations}
\label{eq:dG}
\begin{eqnarray}
\phi_{dGn}&=&+|\phi_G|\\
\phi_{dGs}&=&-|\phi_G|\, .
\end{eqnarray}
\end{subequations}
For the single structures we take $\phi_0=0$ and for the double
gyroids we adjust $\phi_0$ such, that for half the box volume the
density in non-zero.

The functions discussed above can be used as guiding potentials for
nuclear ground state DFT calculations in the Hartree-Fock framework
using the Skyrme functional TOV-min which was developed to reproduce
properties of nuclei and neutron stars~\cite{Erler2013}. Calculations
were done in a periodic box on an equidistant grid using the Sky3D code
from refs.~\cite{Mar15a,Schuetrumpf2018}. All configurations were studied
for proton fractions $X_P=1/2$, $1/3$, and $1/10$ and mean densities
between $\rho=0.04\fm^{-3}$ and $0.08\fm^{-3}$ in steps of
$0.01\fm^{-3}$. The length of the unit cells for the different shapes
have been optimized to minimize the binding energy. Details of the
ground state calculations can be found in~\cite{Schuetrumpf2019}.

\subsection{Neutrino opacities}
\label{sec:neutrinoscatt}

In this subsection, we briefly summarize the formalism of neutrino
scattering on nuclear matter. More complete and detailed descriptions
can be found e.g. in \cite{Horowitz2002,Shen2018,Bedaque2018}.

In a neutron star environment neutrinos can be scattered elastically
and inelastically on the nuclear matter.
The opacity or cross section per volume for
neutral-current scattering on a non-relativistic gas of
neutrons or protons is
\begin{align}
  \frac{1}{V}\frac{\mathrm{d}^2\sigma(E_\nu)}{\mathrm{d}(\cos\theta)\,\mathrm{d}q_0}=&
                                                                                         \frac{G_F^2}{4\pi^2}n
                                                                                         (E_\nu-q_0)^2\left[c_V^2(1+\cos\theta)S_V(q_0,\vec{q})\right.\nonumber\\
                                                                                       &\left.+c_A^2(3-\cos\theta)S_A(q_0,\vec{q})\right]\, ,\label{eq:crosssection1}
\end{align}
where $G_F$ is the Fermi coupling constant, $E_\nu$ is the incoming
neutrino energy, $q_0$ is the energy transfer, $\vec{q}$ is the
momentum transfer and $\theta$ is the scattering angle. $n$ is the
mean neutron or proton number density. The vector and axial coupling
constants $c_V$ and $c_A$ are listed in Tab.~\ref{tab:coupconst}.

\begin{table}[t]
    \centering
    \begin{ruledtabular}
    \renewcommand{\arraystretch}{1.2}
    \begin{tabular}{ccc}
    Reaction&$c_V$&$c_A$  \\\hline
    $\nu p\rightarrow\nu p$&$1/2-2\sin^2\theta_W$&$g_A/2$\\
    $\nu n\rightarrow\nu n$&$-1/2$&$-g_A/2$\\
    \end{tabular}
    \end{ruledtabular}
    \caption{Coupling constants taken from \cite{Horowitz2002}. We
      take the values $\sin^2\theta_W=0.23$ and $g_A=1.26$.} 
    \label{tab:coupconst}
\end{table}

The only terms in Eq.~(\ref{eq:crosssection1}) which depend on nuclear
structure are the dynamic structure factors~\cite{Shen2018,Bedaque2018}
\begin{subequations}
\label{eq:dynsfacs}
\begin{align}
S_V(q_0,\vec{q})&= \frac{1}{2\pi n}
\int\mathrm{d}t\,e^{\I q_0t}
\langle \Phi_0|\hat{\rho}(t,\vec{q})\hat{\rho}(0,-\vec{q})|\Phi_0\rangle\, ,\\
S_A(q_0,\vec{q})&= \frac{2}{3\pi n}
\int\mathrm{d}t\, e^{\I q_0t}
\langle \Phi_0|\hat{\vec{s}}(t,\vec{q})\hat{\vec{s}}(0,-\vec{q)}|\Phi_0\rangle \, .
\end{align}
\end{subequations}
$|\Phi_0\rangle$ is the ground state wave function and 
\begin{subequations}
  \begin{equation}
    \label{eq:rho}
    \hat{\rho}(0,\vec{q}) = \frac{1}{V} \sum_{i=1}^{N} e^{-\I\vec{q}\cdot\vec{r}_i},
  \end{equation}
  \begin{equation}
    \label{eq:srho}
    \hat{\vec{s}}(0,\vec{q}) = \frac{1}{V} \sum_{i=1}^{N} \hat{\vec{s}}_i e^{-\I\vec{q}\cdot\vec{r}_i},
  \end{equation}
\end{subequations}
are the Fourier transforms of density and spin-density operator where $\hat{\vec{s}}_i$ the spin operator acting on the $i$ neutron or proton.
The time dependence of the operators is generated by propagating the operators (rather than the wave  function) with the time-evolution operator of the DFT mean field. The structure factors are normalized such that
$S(q\to\infty)=1$.

The computation of the dynamic structure factors
Eq.~(\ref{eq:dynsfacs}) requires either a fully time-dependent
simulation or to resolve the full excitation spectrum of the
system. However, for neutrino energies larger than the typical nuclear
excitation energies of a few MeV, one can integrate out the dependence
on $q_0$ in Eq.~(\ref{eq:crosssection1}) yielding~\cite{Bedaque2018}
\begin{align}
    \frac{1}{V}\frac{\mathrm{d}\sigma(E_\nu)}{\mathrm{d}(\cos\theta)}=&
    \frac{G_F^2}{4\pi^2}n E_\nu^2\left[c_V^2(1+\cos\theta)S_V(\vec{q})\right.\nonumber\\
    &+ \left.c_A^2(3-\cos\theta)S_A(\vec{q})\right]\,,\label{eq:dcross}
\end{align}
where
\begin{equation}
  \label{eq:statics}
    S_{V/A}(\vec{q})
    =
   \int\mathrm{d}q_0\,S_{V/A}(q_0,\vec{q})
\end{equation}
are the static structure factors. The gain is that the
$S_{V/A}(\vec{q})$ can be determined solely from the ground state
configuration, which is easier to compute.

The transport opacity is given by
\begin{equation}
    \chi_T= \frac{\sigma_T}{V}=\int \mathrm{d}(\cos\theta)\, \frac{\mathrm{d}\sigma}{V\mathrm{d}(\cos\theta)}(1-\cos\theta)
\end{equation}
If we also average the cross section Eq.~(\ref{eq:dcross}) over the
orientation of the incoming neutrino angle with respect to the pasta
structure, we obtain 
\begin{equation}
  \chi_T(E_\nu)=\frac{2G_F^2E_\nu^2}{3\pi}n\left(c_V^2\langle S_V(E_\nu)\rangle+5c_A^2\langle S_A(E_\nu)\rangle\right)\label{eq:totcross}
\end{equation}
with
\begin{subequations}
\begin{eqnarray}
\langle S_V(E_\nu)\rangle&=&\frac{3}{4}\int_{-1}^1\mathrm{d}x\, (1-x^2)S_V(q)\\
\langle S_A(E_\nu)\rangle
&=&
\frac{3}{20}\int_{-1}^1\mathrm{d}x\,(1-x)(3-x)S_A(q)
\\
q &=& \sqrt{2E_\nu^2(1-x)}
\end{eqnarray}
\end{subequations}
where $x=\cos\theta$. The normalization ensures that the averaged
structure factors are comparable.  The factors in
Eq.~(\ref{eq:totcross}) emphasize that the axial part is about five
times as strong as the vector part. $S_{V/A}(q)$ stand for the angular
averaged structure factors. The total transport opacity is finally
given as
\begin{equation}
    \chi^{\text{tot}}_T=\frac{1}{\lambda_T}= \chi_{T,p} +\chi_{T,n}
\end{equation}
where $n$ and $p$ label neutrons and protons. The mean neutrino
opacity is defined by averaging over the neutrino spectrum 
\begin{equation}
\langle \chi^{\text{tot}}_T \rangle = \langle 1/\lambda_T \rangle =\int
\chi(E_\nu) f_\nu(E_\nu,T)\, E_\nu^2\mathrm{d}E_\nu \, , 
\end{equation}
where $f_\nu(E_\nu,T)$ is the neutrino distribution function for which
we take a normalized Boltzmann distribution. In the above discussion
we have considered neutral-current scattering but can be easily
generalized to consider charged-current (anti)neutrino absorption
reactions.

\subsection{Static structure factors}

As we have seen in the section \ref{sec:neutrinoscatt}, the structure
factors are the nuclear input to the scattering cross sections. The
static structure factor of eq.~\eqref{eq:statics} can be expressed
as~\cite{Horowitz2004}: 

\begin{equation}
  \label{eq:stoF}
  S(\vec{q}) = \frac{1}{N} \sum_n |\langle \Phi_n |\hat{F}(\vec{q})
  |\Phi_0 \rangle|^2,
\end{equation}
with the sum running over excited states $|\Phi_n\rangle$. 
$\hat{F}(\vec{q})$ is the form factor operator that in isospin
formalism can be expressed as
\begin{subequations}
\begin{equation}
 \hat{F}_\mathrm{tt'}(\bm{q})=\sum_{i=1}^N e^{\mathrm{i}\vec{q}\cdot\vec{r_i}}\hat{T}_\mathrm{tt'}\quad,
\label{eq:formf}
\end{equation}
with the isospin selectors
\begin{align}
  \hat{T}_{nn}=\hat{T}_n&=\frac{1+\hat{\vec{\tau}}_0}{2}\quad,&
  \hat{T}_{pp}=\hat{T}_p&=\frac{1-\hat{\vec{\tau}}_0}{2}\quad,\nonumber\\
  \hat{T}_{np}&=\hat{\vec{\tau}}_+\quad,&
  \hat{T}_{pn}&=\hat{\vec{\tau}}_-\quad,
\end{align}
\end{subequations}
where the $\vec{\tau}$ are standard isospin matrices.  The form factor
operators are not hermitean, but obey the relation
$\hat{F}_\mathrm{tt'}^\dagger(\bm{q})=\hat{F}_{\mathrm{t't}}(-\vec{q})$.
We will abbreviate the diagonal part often as
$\hat{F}_\mathrm{tt}=\hat{F}_\mathrm{t}$.

The static structure factors $S_{V/A}(\vec{q})$ can be decomposed into
two parts, elastic and inelastic structure factors.

\subsubsection{Elastic Structure Factor}

The elastic vector structure factor includes only the contribution of
the ground state in eq~\eqref{eq:stoF} and for a single unit cell it
reads
\begin{equation}
  S_{\mathrm{el},t}^{(0)}(\bm{q})
  =\frac{1}{N_t}\left|\langle\Phi_0|\hat{F}_t(\vec{q})|\Phi_0\rangle\right|^2
  \quad,
\label{eq:elastelem}
\end{equation}
Since our calculations respect time-reversal symmetry the spin density
is zero for the ground-state and hence the axial elastic structure
factor is zero. Because our calculations are done for one elementary cell, we need to take the limit to infinite volume of the
expression above. As the form factor scales with $N_t$, the number of
particles of type $t$ in the elementary cell, the height of the
elastic structure factor peaks still scale with $N_t$, however the
peaks become more narrow with increasing number of particles. Special
attention has to be paid when performing the limit for an infinite
system when the unit cell of the pasta configuration is repeated
periodically. We consider here configurations which are perfectly
periodic in all three space dimensions. Thus the form factor is only
non-zero at the $\vec{k}$-space points $\vec{k}_\mu=2\pi \vec{\mu}/L$,
with $\vec{\mu}\in\mathbb{Z}^3$. Then the elastic structure factor becomes
\begin{equation}
      S_{\mathrm{el},t}(\bm{q})=
      (2\pi)^3 n_t\sum_\mu \delta^3 (\vec{q}-\vec{k}_\mu)
      \frac{1}{N_t}S_{\mathrm{el},t}^{(0)}(\vec{k}_\mu)
\label{eq:Sel}
\end{equation}
where $\vec{k}_\mu$ are the reciprocal lattice vectors,
$S_{\mathrm{el},t}^{(0)}$ is the structure factor
Eq.~(\ref{eq:elastelem}) for one elementary cell. In the infinite
limit, the elastic structure factor does not scale with particle
number, as in a finite system, but with the mean number density $n_t$.

The angular averaged structure factor, which is used to determine
the neutrino opacity, becomes then
\begin{equation}
        S_{\mathrm{el},t}(q)=
      2\pi^2n_t\sum_\mu \frac{\delta (q-k_\mu)}{q^2}
      \frac{1}{N_t^2}|\langle\Phi_0|\hat{F}_t(\vec{k}_\mu)|\Phi_0\rangle|^2\, .
\end{equation}

\subsubsection{Inelastic Structure Factor}

The inelastic structure factor for a unit cell is determined as the
elastic one in Eq.~\eqref{eq:elastelem}, however including excitations
from the ground state to an excited state $\Phi_n$.
\begin{equation}
  S_{\mathrm{inel},tt'}^{(0)}(\bm{q})=
  \frac{1}{N_t}\sum_{n>0}
  \langle\Phi_0|\hat{F}_{tt'}(-\bm{q})|\Phi_n\rangle
  \langle\Phi_n|\hat{F}_{t't}(\bm{q})|\Phi_0\rangle
  \,.
\label{eq:Sinel}
\end{equation}
For the cell sizes and particle numbers considered here, finite size
effects on the inelastic structure factor are small and our
calculations represent already a very good approximation to the
infinite system inelastic structure factor by taking
$S_{\mathrm{inel},tt'}(\bm{q})=S_{\mathrm{inel},tt'}^{(0)}(\bm{q})$.

If the spin-orbit interaction and thus the spin-mixing is small, the
axial inelastic part can be approximated through the vector inelastic
part. We have checked that this is a good approximation in our
calculations and will be used in the following. In summary, we have
\begin{align}
  S_{V,\mathrm{el},t}&=S_{\mathrm{el},t}\quad,&
  S_{V,\mathrm{inel},tt'}&=S_{\mathrm{inel},tt'},\nonumber\\
  S_{A,\mathrm{el},t}&=0\quad,&
  S_{A,\mathrm{inel},tt'}&\approx S_{\mathrm{inel},tt'}.
\end{align}

\subsubsection{Generalization to Finite Temperature}

Both definitions of the structure factors, Eqs.~(\ref{eq:Sel}) and
(\ref{eq:Sinel}), apply to zero temperature. At finite temperature,
the system is described by a density operator
$|\Phi_0\rangle\langle\Phi_0|\rightarrow\hat{D}_0$ for the thermal
ground state. For a Slater determinant state $|\Phi_0\rangle$, $\hat{D}_0$
can be expressed as
\begin{equation}
  \hat{D}_0=
  \prod_\alpha
  \frac{\exp\left(-\hat{a}_\alpha^\dagger\hat{a}_\alpha^{\mbox{}}\varepsilon_\alpha/T\right)}
       {1+e^{-\varepsilon_\alpha/T}}
\end{equation}
where $T$ is the temperature and $\alpha$ runs over a complete set of
single-particle states $\alpha$. These are eigen states of the
mean-field Hamiltonian with single-particle energy
$\varepsilon_\alpha$.  The detailed expressions for expectation values
with $\hat{D}_0$ in terms of single-particle wave functions
$\varphi_\alpha$ can be evaluated with standard methods of many-body
theory~\cite{Bal86aR}.
The static structure factors at finite $T$ for a unit cell  in
Eq.~(\ref{eq:Sel}) becomes
\begin{subequations}
\label{eq:structurefactors2}
\begin{align}
  S_{\mathrm{el},t}^{(0)}(\bm{q})&=\frac{1}{N_t}
   \mbox{tr}\{\hat{F}_t(-\vec{q})\hat{D}_0\hat{F}_t(\vec{q})\hat{D}_0\}
\nonumber\\
  &=
  \frac{1}{N_t}\left|\sum_\alpha w_\alpha\langle\varphi_\alpha
  |e^{\mathrm{i}\vec{q}\cdot\vec{r}}\hat{T}_\mathrm{t}
  |\varphi_\alpha\rangle\right|^2\quad,
\\
  S_{\mathrm{inel},tt'}^{(0)}(\bm{q})
  &=
  \frac{1}{N_t}\Big[\mbox{tr}\{\hat{F}_{tt'}(-\bm{q})
    (1-\hat{D}_0)\hat{F}_{t't}(\bm{q})\hat{D}_0\}
\nonumber\\
  &=
  1-\frac{1}{N_t}\!\sum_{\alpha,\beta} w_\alpha w_\beta
  \left|\langle\varphi_\alpha|e^{\mathrm{i}\vec{q}\cdot\vec{r}}
  \hat{T}_\mathrm{tt'}|\varphi_\beta\rangle\right|^2\,,
\\
  w_\alpha&=\frac{1}{1+e^{-\varepsilon_\alpha/T}}\quad,
\end{align}
\end{subequations}
where $w_\alpha$ is the thermal occupation probability of the
single-particle state $\varphi_\alpha$.

\section{Results}\label{sec:results}
\subsection{Elastic structure factor}

\begin{figure}[!t]
    \centering
    \includegraphics[width=\columnwidth]{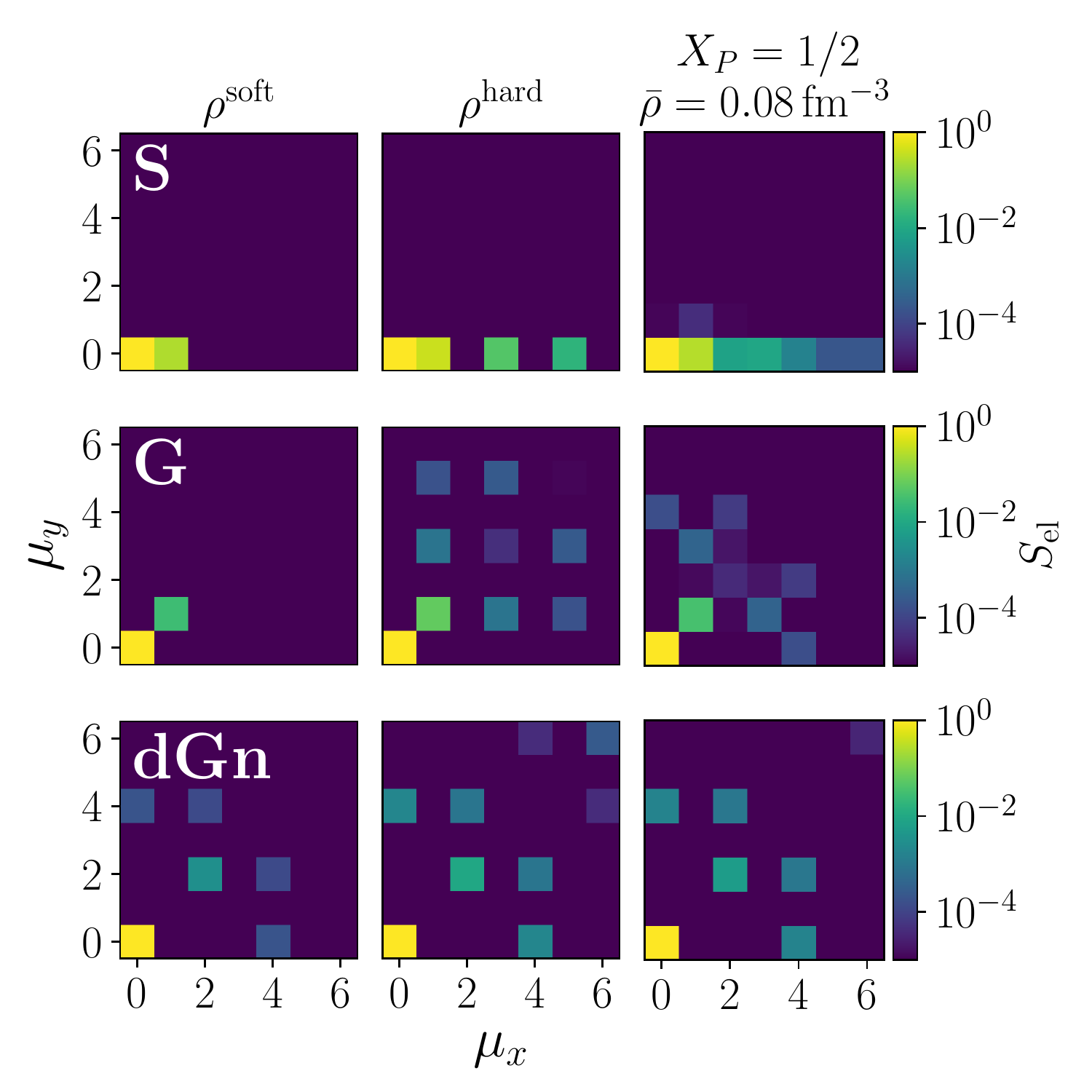}
    \caption{First components of the elastic structure function in the
      $x$-$y$-plane. The $\mu_x$ and $\mu_y$ label the Fourier bins in $x$-
      and $y$-direction. Upper row shows slab, middle Gyroid, and lower
      double Gyroid network-like configurations. First column displays
      the smooth model (\ref{eq:soft}), second column is the hard
      model (\ref{eq:hard}), and third column is actual pasta
      configuration (neutrons) from DFT calculations with proton
      fraction $X_P=1/2$ and average density
      $\bar{\rho}=0.08\fm^{-3}$.}
    \label{fig:form2d}
\end{figure}

The elastic structure factor is proportional to the normalized
absolute square of the Fourier transform of the particle density.  The model structures
(\ref{eq:nodal_approx}) indicate that the different geometries should
be distinguished by marked difference in the distribution of these
Fourier components. This surely holds for 
three-dimensional distributions before angular averaging.  This is
shown in Fig.~\ref{fig:form2d} which shows the first 6 Fourier
coefficients of the first quadrant in the $\mu_x$-$\mu_y$-plane of the
elastic structure function for S, G, and dGn geometries.  In the left
column the densities are modeled after the ``soft'' approximation
Eq.~(\ref{eq:soft}),the second column shows the result for the
``hard'' approximation Eq.~(\ref{eq:hard}). Finally, the third column
shows results from realistic, self-consistent 3D simulations of the
pasta configurations at $T=0$.

The soft approximation shows already the leading peaks of the elastic
structure function.  In fact, the distribution and predominance of
Fourier components can be read off from the analytical structures in
Eqs. (\ref{eq:nodal_approx}).  This is straight-forward for the single
configurations and a bit more involved for the double configurations
because they contain the absolute value
(Eqs.~(\ref{eq:dG})). 
\begin{table}[!t]
    \centering
    \begin{ruledtabular}
    \begin{tabular}{llrr}
    shape               & $|\vec{\mu}|$ & $\vec{\mu}$  &\\\hline
    S                   & 1           &($\pm$1,0,0)            &\\
    P                   & 1           &($\pm$1,0,0)            & and permutations\\
    G                   & $\sqrt{2}$  &($\pm$1,$\pm$1,0)       & and permutations\\
    D                   & $\sqrt{3}$  &($\pm$1,$\pm$1,$\pm$1)  &\\
    \multirow{2}{*}{dG} & $\sqrt{6}$  &($\pm$2,$\pm$1,$\pm$1)  & and permutations\\
                        & $\sqrt{8}$  &($\pm$2,$\pm$2,0)       & and permutations
    \end{tabular}
    \end{ruledtabular}
    \caption{First dominant peak(s) in the elastic structure factor $S_\mathrm{el}$ for the different pasta configurations.}
    \label{tab:Min_surf}
\end{table}
Complementing Fig.~\ref{fig:form2d}, we list in
Tab.~\ref{tab:Min_surf} the first dominant peaks for all
configurations considered in this work, and here for all three
directions in $\mathbf{k}$ space.

The hard distribution, middle column in Fig.~\ref{fig:form2d}, shows
already secondary peaks, because the steep transition at the surface
enhances higher Fourier components. Note that the single
configurations in this model occupy only peaks with odd $\mu_i$ (except
for $\vec{\mu}=0$) while double configurations can occupy all
$\vec{\mu}$. The self-consistent DFT pasta configurations (right panels)
occupy all Fourier coefficients. For the slab shape ``S'' almost
exclusively the coefficients in x-direction are occupied, because the
configuration is homogeneous in the other two directions. One can spot
only very small disturbances into other directions due to spatial
fluctuations in the calculations. The Gyroid as well as the double
Gyroid is very close to the hard configurations.

\begin{figure}[!t]
    \centering
    \includegraphics[width=\columnwidth]{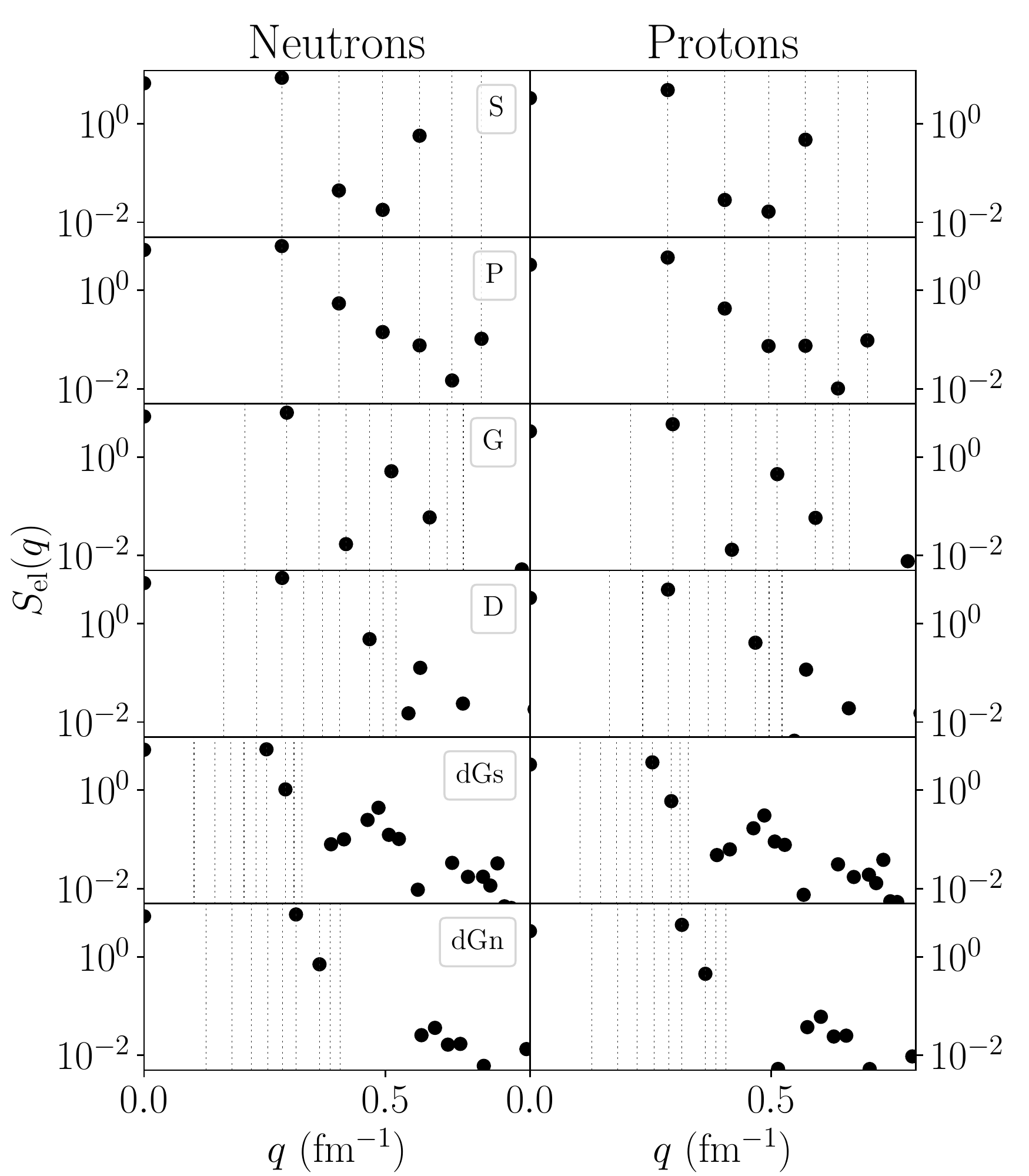}
    \caption{Angular averaged elastic structure factor for all studied
      pasta configurations for $\rho=0.04\fm^{-3}$ and $X_P=1/3$. Left
      panels: neutrons; right panels: protons. Additionally, the first
      possible bins, up to $\vec{\mu}^2=10$ are shown.} 
    \label{fig:Sel}
\end{figure}

Pasta matter in the stellar environment will appear as isotropic
distribution of finite fragments of nuclear matter. This suggests that
we are now looking at angular averaged distributions which are radial
distributions in momentum space.

Fig.~\ref{fig:Sel} shows the elastic structure factor as function of
the momentum transfer $q$ for $\rho=0.04\fm^{-3}$ and proton fraction
$X_P=1/3$. It is striking that the first peak appears for all
different configurations at about the same value $q=0.31\fm^{-1}$,
although the peaks are at very different bins in Fourier space
(compare to Fig.~\ref{fig:form2d} and Tab.~\ref{tab:Min_surf}). For
comparison we also show the first few possible
$q$-bins according to the box
lengths $q_\mathrm{bin}=\frac{2\pi}{L}|\vec{\mu}|$. It seems that
the peak number in Fourier space is counter weighted by the cell size
of the geometry delivering eventually all first maxima at about the
same $q$ value. However, the further evolution of peaks with their
fine structure is sensitive to the exact pasta configuration.

Proton and neutron peak structures are very similar, also for a proton
fraction of $X_P=1/10$. The most important difference is that the
proton distribution decays more slowly with large $q$, because the
protons are more localized for $X_P<1/2$ while the neutrons form a
neutron skin, sometimes even a finite background gas.

\begin{figure}[!t]
    \centering
    \includegraphics[width=\columnwidth]{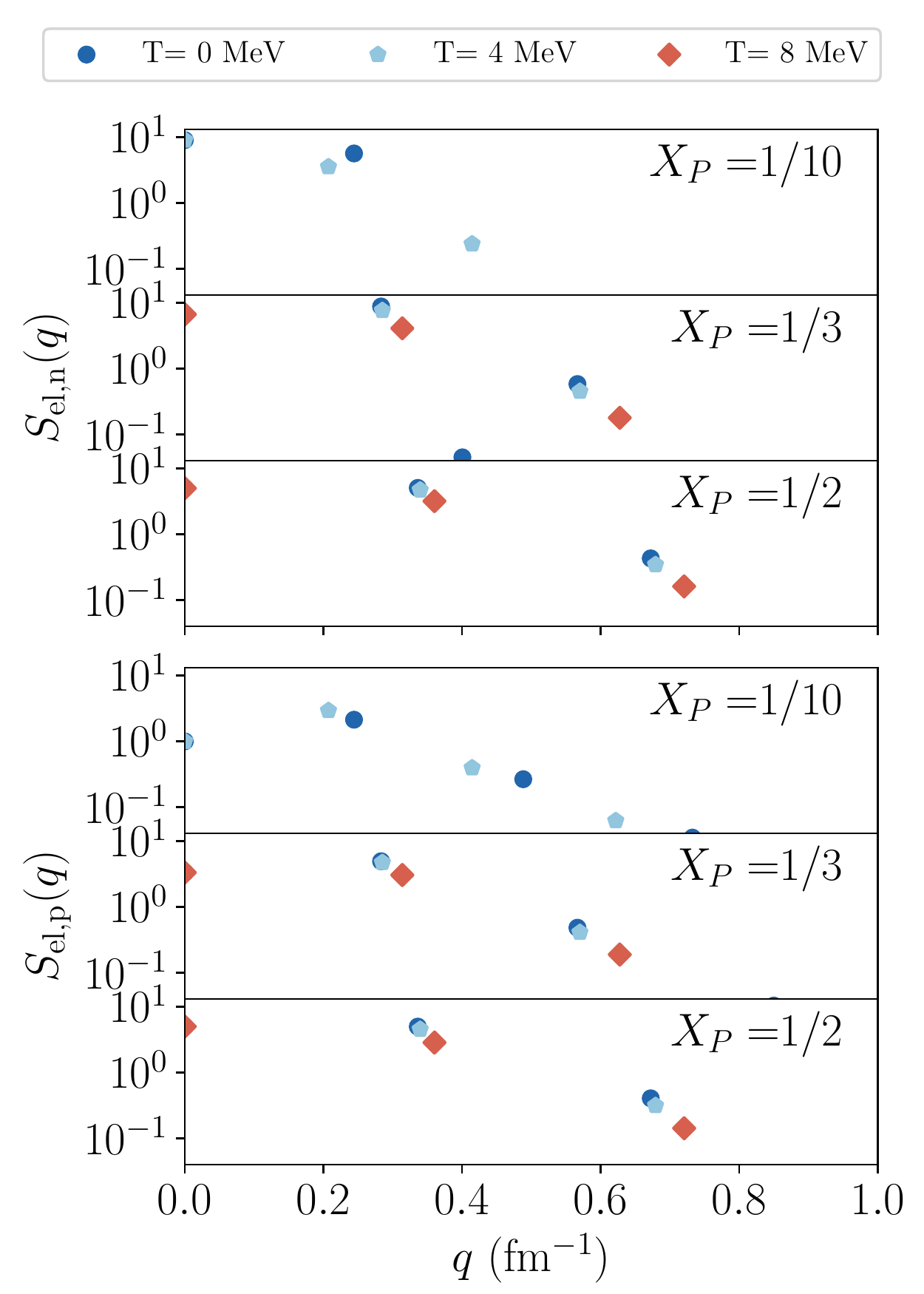}
    \caption{Elastic structure factor for the slab shape at $\rho=0.04\fm^{-3}$ for different temperatures and proton fractions.}
    \label{fig:SelT}
\end{figure}

We have also checked the impact of proton content $X_P$ and of
temperature $T$.  It turns out that the trends are very similar in all
geometries.  Thus we restrict the analysis here to the slab geometry.
Figure~\ref{fig:SelT} shows the elastic structure factors for the slab
shape at different temperatures and proton fractions. The position of
the first peak moves with increasing proton fraction and temperature. Its trend
is closely related to box size, the smaller the box the higher the $q$
of the peak. The preferred box lengths for the slabs were studied in a
previous paper~\cite{Schuetrumpf2019}. They become larger for smaller
proton fractions. The box lengths increase for $T=2\MeV$ and decrease
for higher temperatures.

Since the slab shape is homogeneous in two directions and has only
non-zero contributions in one direction. The peak structure is
relatively simple being equidistant in $q$. The only exception is the
slab at $T=0$ and $X_P=1/3$, where small additional peaks show up
which happens because there are some fluctuations into the other two
directions.

The single peaks are decreasing in magnitude for larger $q$ values
which is typical for form factors. This trend becomes stronger for
smaller proton fraction and higher temperature. This is due to
stronger smoothing of the surfaces and it is especially visible for
neutrons, which form a neutron background gas for high temperatures or
small proton fractions.

\subsection{Inelastic structure factor}

\begin{figure}[!t]
    \centering
    \includegraphics[width=0.8\columnwidth]{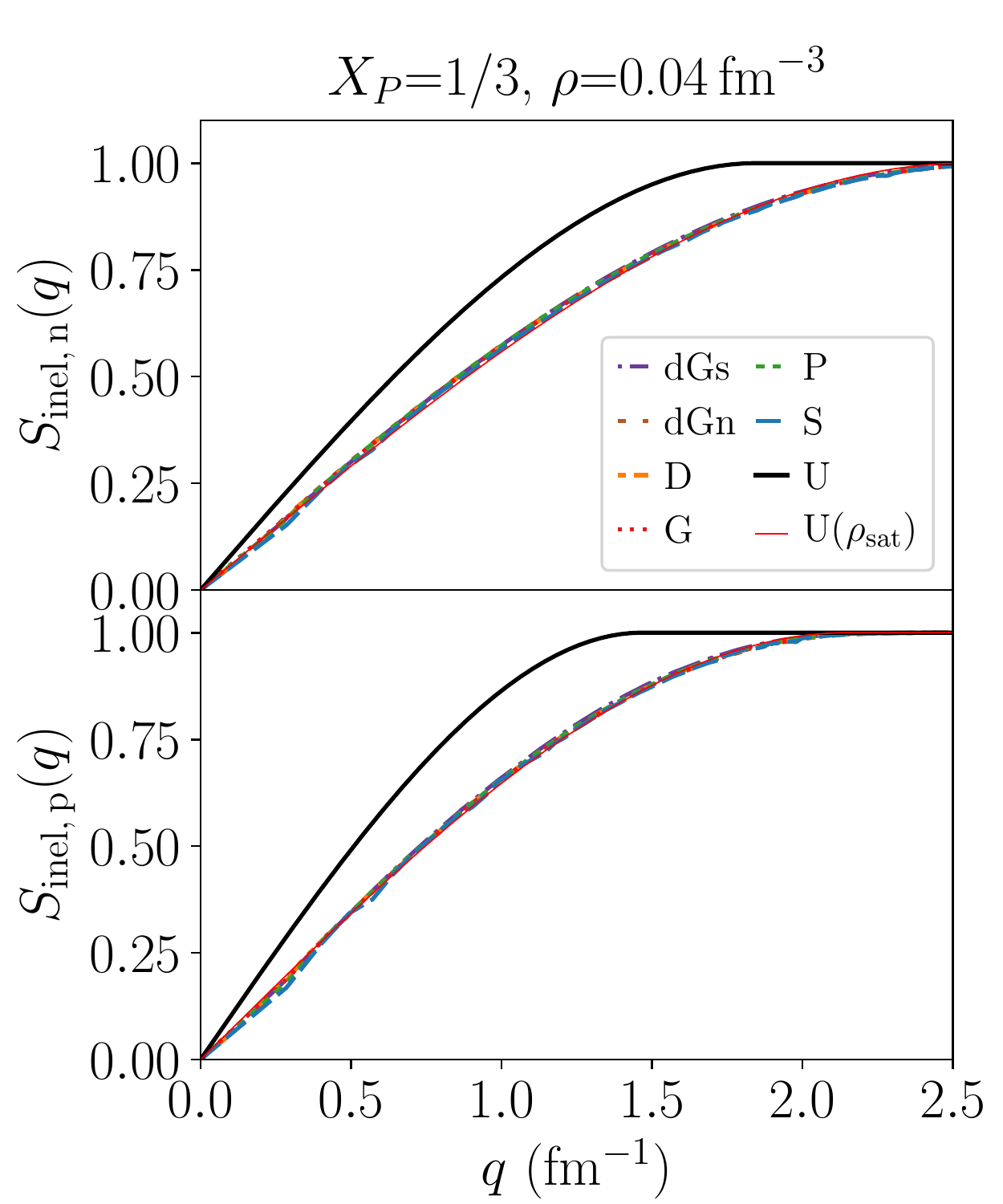}
    \caption{Inelastic structure factor for all pasta configurations
      (dashed lines) and uniform matter with the same mean density
      (thick solid line) as well as uniform matter with saturation
      density in the pasta solid phase (thin solid line). Upper panel:
      neutrons; lower panel: protons.}
    \label{fig:Sinel}
\end{figure}

In contrast to the elastic structure factor, the inelastic structure
factor is a smooth function also for strictly periodic systems,
because it is sensitive to the detailed wave functions and these wave
functions are only quasi-periodic (Bloch states).
Figure~\ref{fig:Sinel} shows the inelastic proton and neutron
structure factors for all studied pasta configurations for $X_P=1/3$
and $\rho=0.04\fm^{-3}$ compared with the one for uniform matter.
First, it turns out that the structure factors are identical for all
pasta configurations, except for small deviation stemming from
finite-size effects. However, compared to uniform matter at the same
mean density (heavy black line labelled U), the inelastic structure
factors are reduced.  However, considering uniform matter having the
density $\rho_{\mathrm{sat}}$ which is found inside the high density
pasta phase (small red line denoted U($\rho_\mathrm{sat}$)), we find
practically the same pattern as for the pasta structures.  There is a
one-to-one correspondence to saturation density
$\rho_{\mathrm{sat}}$. The reason is that the inelastic structure
factor in DFT only reflects the Pauli correlation in the filled
volumes of the pasta configurations.  Higher correlations are not
included. Real two-particle correlations can be resolved by other
methods like molecular dynamics (MD)
\cite{Horowitz2004,Horowitz20042,Nandi2018}. However, in MD
simulations the Pauli exclusion principle is not strictly reproduced,
as a phenomenological potential is implemented to simulate the
effect. To include both, Pauli effects as well as higher many-particle
correlations, it would be necessary to perform RPA
calculations~\cite{Pastore2015,Bedaque2018,Dzhioev2018} or use Monte
Carlo methods~\cite{Shen2018}. 


\begin{figure}[!t]
    \centering
    \includegraphics[width=\columnwidth]{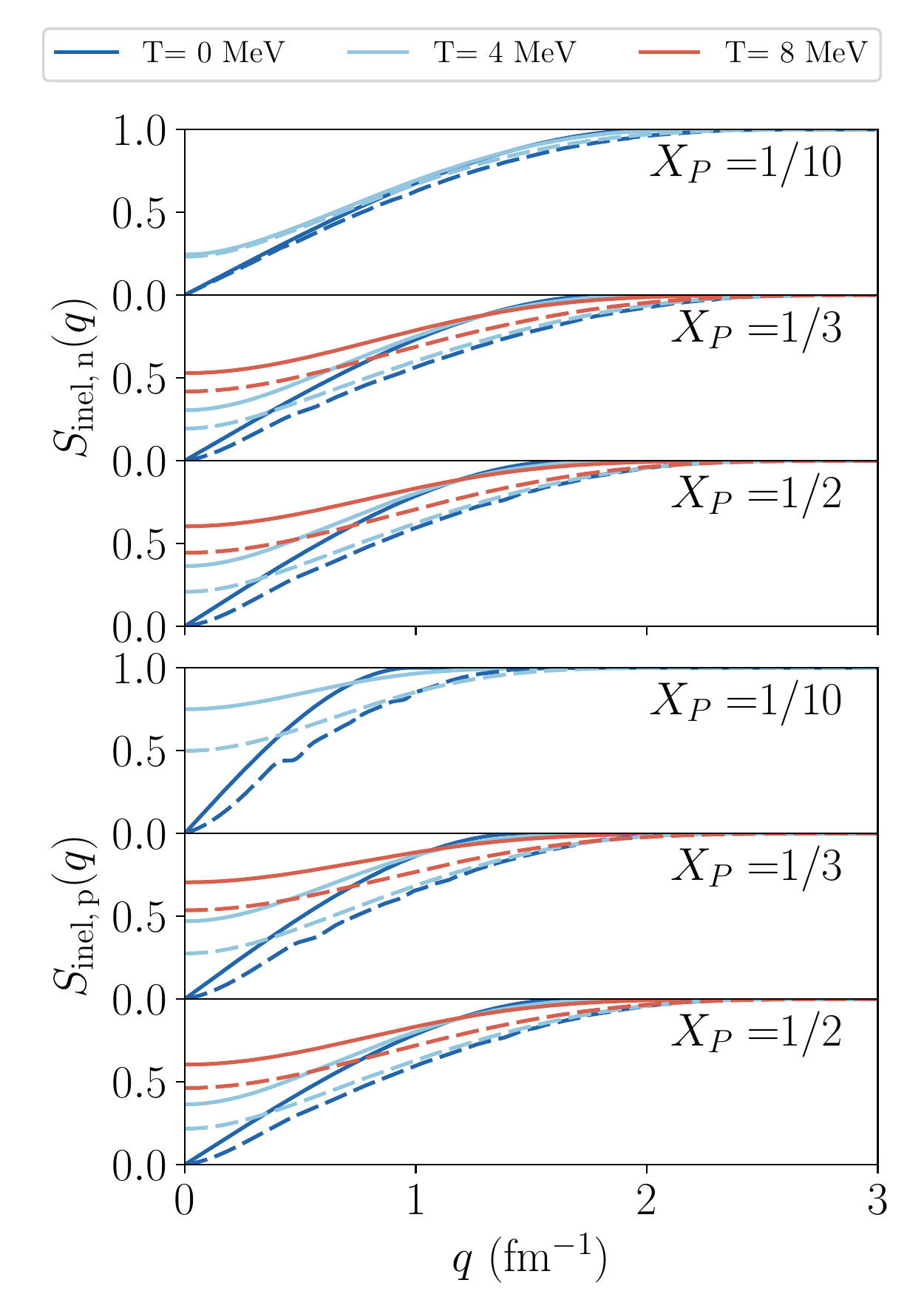}
    \caption{Inelastic structure factor for the slab shape (dashed lines) for  $\rho=0.04\fm^{-3}$ at different temperatures and proton fractions in comparison to uniform matter (solid lines).}
    \label{fig:SinelT}
\end{figure}

Fig.~\ref{fig:SinelT} shows the neutral-current, inelastic structure
factors for varying temperatures and proton fractions.  The simulated
results shows occasionally small wiggles in the proton structure
factors at zero temperature. These are caused by fluctuations from
finite particle numbers in the box which are unimportant for our
discussion. The $T=0$ results are the same shown before where, again,
the slab results are smaller than the trend from homogeneous matter
(when compared at same average density).  This difference decreases
with increasing temperature, because the simulated slab acquires
increasingly softer surface and slowly approaches homogeneous matter.
For $X_P=1/10$ the structure factor of neutrons for the slab
is already very close to uniform matter due to the large neutron
background density. In contrast, the proton structure factors for
slabs and uniform matter differ significantly.  The trend of the
inelastic structure factors at $\vec{q}=0$ with temperature deserves
a comment. The value $S_{\mathrm{inel},tt}(0)$ starts at zero for
$T=0$ which expresses the Pauli principle between two like nucleons.
The value increase with temperature, because higher momenta in phase
space are becoming occupied and lower momenta in phase space become 
unoccupied such that the Pauli blocking is reduced.
The effect is independent of the spatial profile of matter.

\begin{figure}[!t]
    \centering
    \includegraphics[width=\columnwidth]{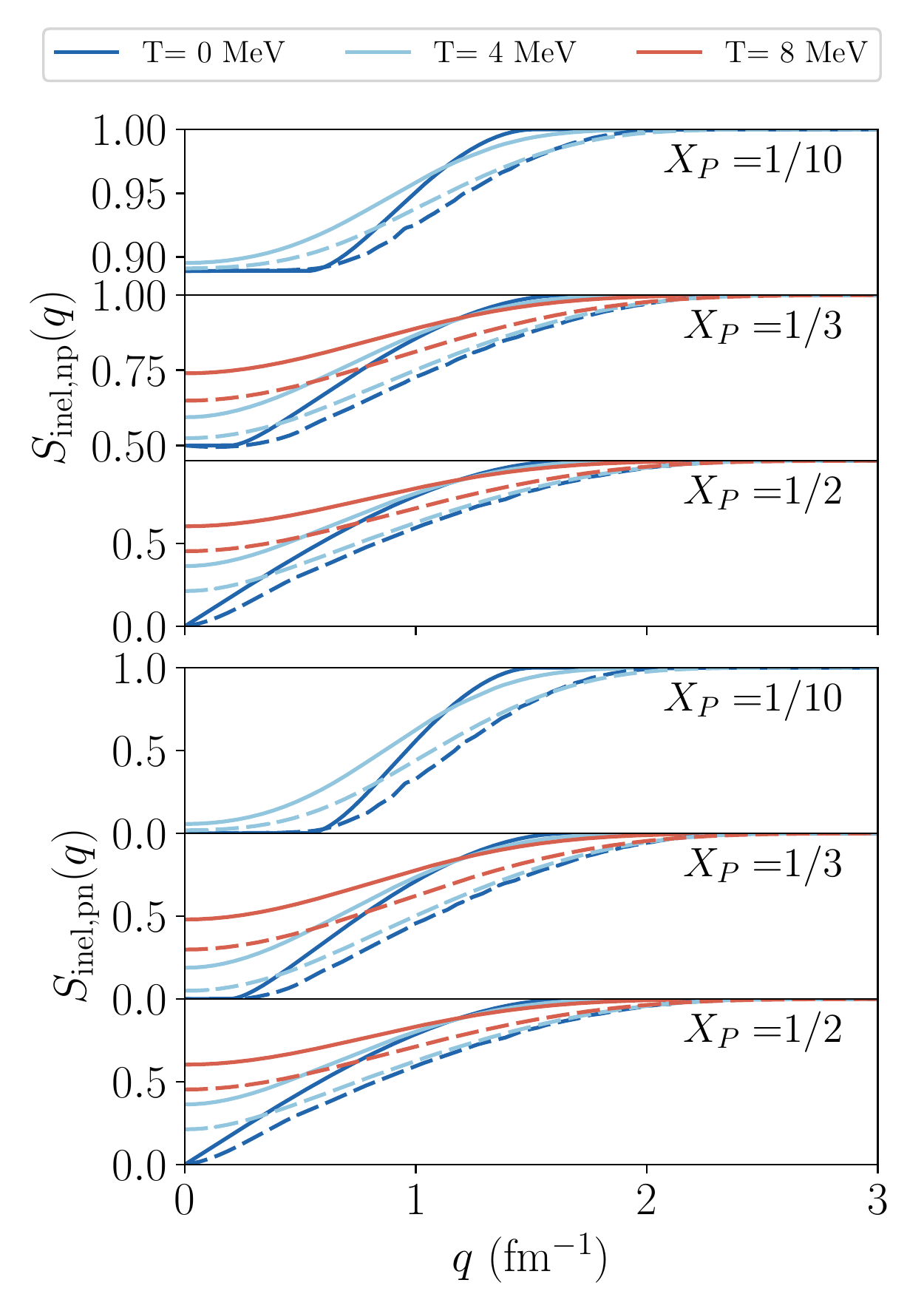}
    \caption{Charged-current inelastic structure factors for the slab
      shape (dashed lines) for $\rho=0.04\fm^{-3}$, different
      temperatures and different proton fractions in comparison to
      uniform matter (solid lines).}
    \label{fig:Sinelc}
\end{figure}

Fig.~\ref{fig:Sinelc} shows the charged-current structure factors. We
see that the differences of the slab to uniform matter are comparable
to those for the neutral-current structure factors. Though, the
overall behavior is a bit different. While for $X_P<1/2$ for
$p\rightarrow n$ reactions the structure factor starts to increase for
higher values of $q$, the structure factor for $n\rightarrow p$ starts
at $q=0$ with a non-zero value and increases with a lower slope. At
low temperatures, again, the difference between homogeneous matter and
pasta is more pronounced.

\subsection{Impact on neutrino opacity}

\begin{figure}[htb]
    \centering
    \includegraphics[width=\columnwidth]{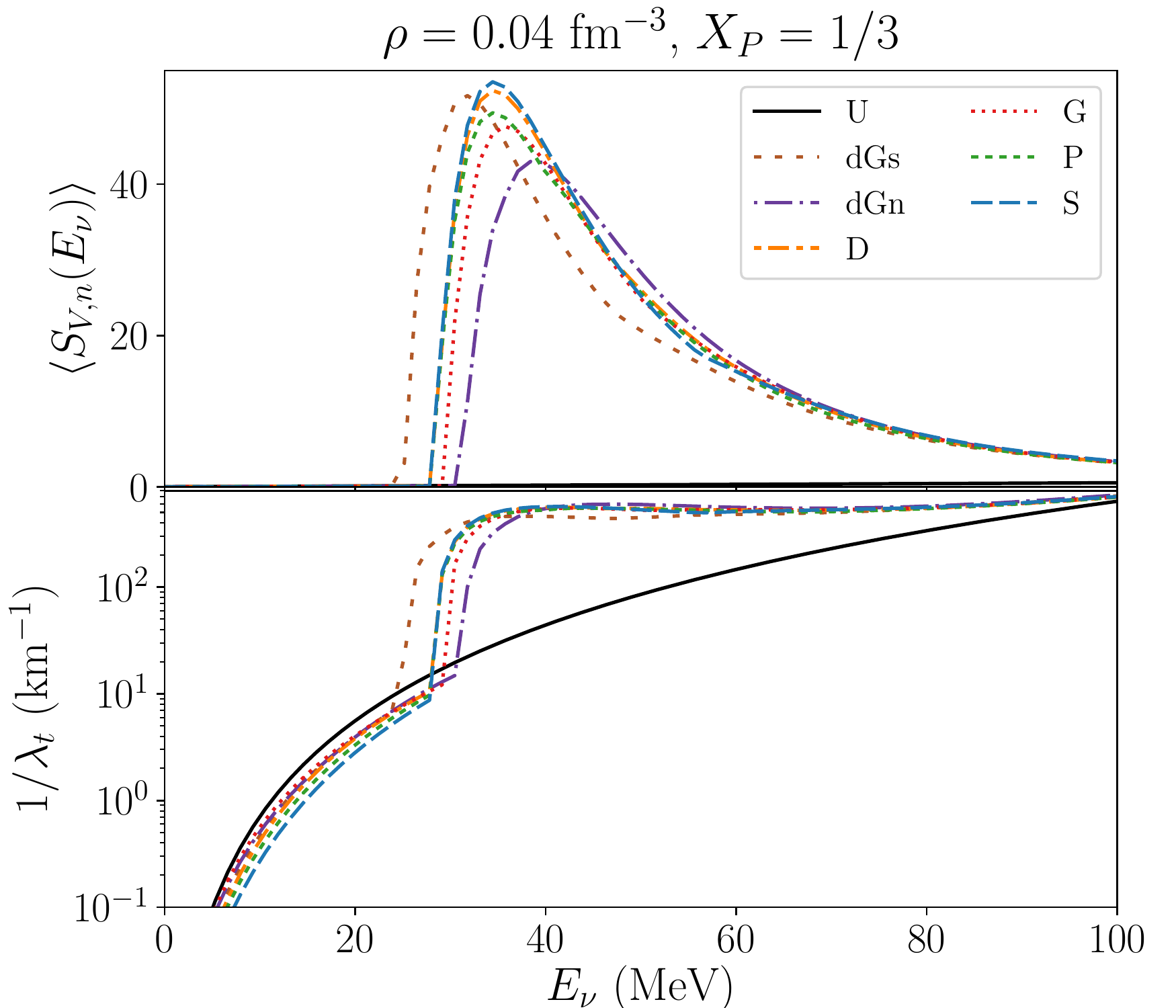}
    \caption{Averaged vector static structure factors for neutrons (upper panel) and opacity (lower panel) for $\rho=0.04 \fm^{-3}$ and $X_P=1/3$ for different pasta configurations and uniform matter.}
    \label{fig:opa_0.3-0.04}
\end{figure}
Finally, we are going to estimate the impact of nuclear pasta matter
on the neutrino opacity. To that end we only consider neutral-current
reactions, because, as we have seen in the previous sections, pasta
matter has only a marginal influence on charged-current reactions,
because there is no elastic scattering channel.

The upper panel of Fig.~\ref{fig:opa_0.3-0.04} shows the averaged
static vector structure factors for neutrons for $\rho=0.04 \fm^{-3}$
and $X_P=1/3$. This structure factor has the most impact on opacity,
because it contains the elastic part. The vector static structure
factors for protons also contains an elastic contribution, however it
is strongly suppressed through the small coupling constant
$c_{V,p}$. The different pasta configurations all show peaks at
approximately the same position and same heights. Only the double
Gyroid configurations deviate slightly.

The lower panel of Fig.~\ref{fig:opa_0.3-0.04} shows the resulting
opacity. While for low neutrino energies pasta matter slightly
reduces the opacity, it enhances it for neutrino energies larger
than 25 to 30 MeV. Note that the impact seems to be lower than in
Ref. \cite{Horowitz2004}, because we also take into account the
contribution from the axial current.

\begin{figure}[htb]
    \centering
    \includegraphics[width=\columnwidth]{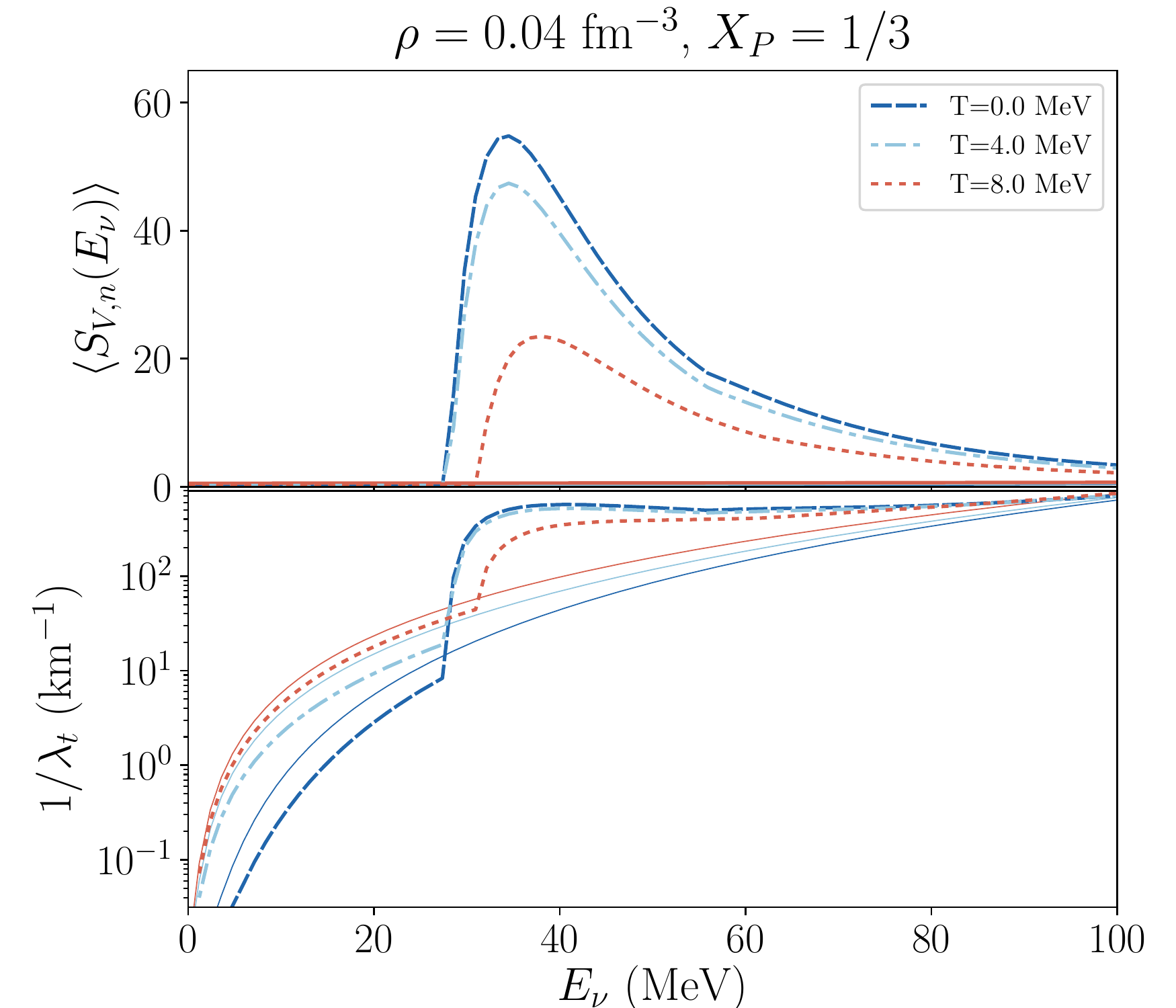}
    \caption{Same as Fig.~\ref{fig:opa_0.3-0.04} but only for the pasta slab configuration (dashed) and for uniform matter (solid) and varying temperatures.}
    \label{fig:opa_0.3-0.04_T}
\end{figure}

Figure~\ref{fig:opa_0.3-0.04_T} shows the same quantities as
Fig.~\ref{fig:opa_0.3-0.04} but only for slab as representative of all
pasta configurations and uniform matter, both for a series of
different temperatures. Slab configurations show a pronounced peak in
the structure factor and subsequently a significant enhancement in
opacity between 30 and 50 MeV.  The effect is largest impact at zero
temperature. 

\begin{figure}[htb]
    \centering
    \includegraphics[width=\columnwidth]{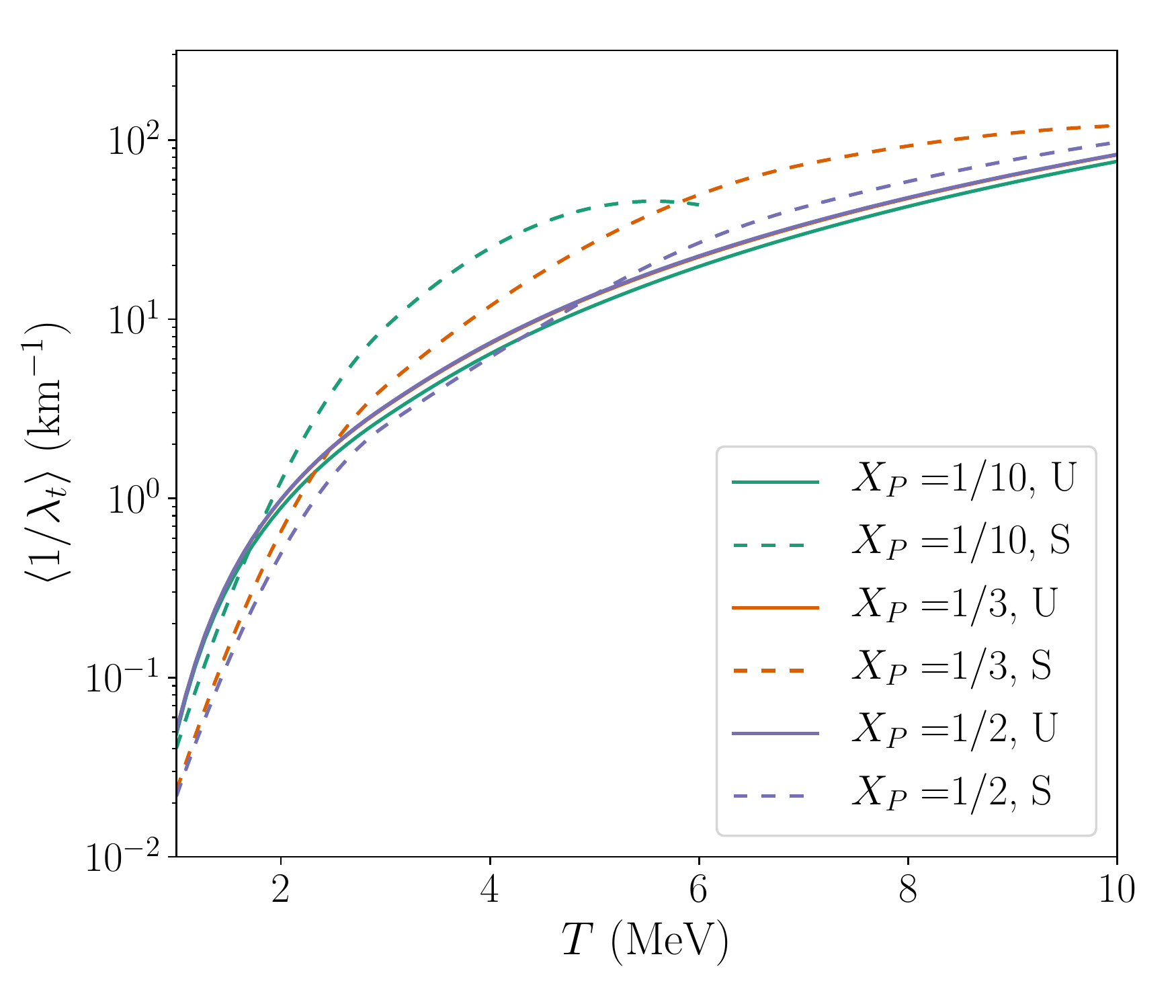}
    \caption{Neutrino opacities for the pasta slab configuration and uniform matter with respect to the temperature for $\rho=0.04\fm^{-3}$ three different proton fractions..}
    \label{fig:opa_mean}
\end{figure}
Fig.~\ref{fig:opa_mean} compares the mean opacity for slab and uniform
matter as functon of temperature for a series of proton
fractions $X_P$. While for symmetric nuclear matter ($X_P=1/2$) the
impact of slab structure is rather small, it becomes larger for smaller proton
fractions. This is because for smaller proton fractions the elastic
peak is shifted to lower neutrino energies (c.f. Fig~\ref{fig:SelT})
and thus has a larger overlap with the neutrino energy distribution
already at lower temperatures. For $X_P=1/10$ the impact of pasta
matter is significant and enhances the opacity by a factor of about
three at a temperature of 4~MeV.

\section{Conclusions}

We calculated the elastic and the inelastic part of the structure
factor for nuclear pasta configurations and for uniform matter. This
was done at a fully quantum mechanical level using nuclear
density-functional theory (DFT). All considered pasta configurations
appear in a cubic, periodic lattice with lattice spacing (box length)
depending on the strutture.  The analysis was complemented by a simple
analytical model of the minimal surfaces corresponding to the most
important pasta structures: Slab (S), Primitive (P), Gyroid (G), and
Diamond (D). Unlike uniform matter, the different pasta configurations
produce an elastic structure factor with rich and distinctive pattern.

The elastic structure factor is a function of transfered momentum.
Periodic lattice structure produces distinctive peaks at the
reciprocal vector of the lattice. The pasta configuration determines
box lengths (length of the elementary cell) and distributions of
strength over the peaks.  The emerging pattern depend sensitively on
the pasta configuration. Nonetheless, the first dominant peak shows up
at roughly the same momentum transfer for all configurations.  Changes
in box length, thus lowest momentum bin, are compensated by the fact
that the first dominant peak appears at different bin number for
different copnfigurations.  The similarity of dominant transfered
momentum means that the overall trend for the elastic structure is
similar for all configurations. Only details change with the changing
configuration.

There are unique trends with changing proton fraction and temperature.
Lower proton content shifts the peaks to lower momentum transfer $q$.
High temperatures $T>4\MeV$ shift the peaks to higher momentum
transfer $q$ and the peaks decrease then more rapidly with higher $q$
because the matter is smoothed and approaches uniform matter.

The inelastic structure factor is found to be independent of the
actual pasta configuration and it is same as uniform matter at a
density which corresponds to the saturation density in the filled
regions of pasta structures. It is only this saturation density which
counts.  This happens because in DFT the inelastic structure factor
only reflects the Pauli correlations as one can also see from the fact
that the structure factors at temperature zero are exactly zero while
finite temperature which override gradually the Pauli blocking allows
for a non-vanishing inelastic structure factor at $q=0$.

Finally, we computed the neutrino opacities for the given structures
as function of temperatures and proton fraction. The results do not
depend much on the actual structure, but differ significantly from
uniform matter, particularly at low temperature and low proton
fraction. Pasta strutures enhance the neutral-current opacity up to a
factor of four. For charged-current the effect of pasta is minor. 

In summary, nuclear pasta has a noticeable impact on the structure
factors and thus on inelastic and elastic neutrino scattering. While
for inelastic reactions the cross section is reduced, the elastic
coherent scattering increases dramatically. The effect can be of
importance for the cooling of neutron stars as well as for
core-collapse supernova models.

\begin{acknowledgements}
  Computational resources were provided by the Center for Scientific
  Computing (CSC) of the Goethe University Frankfurt.  This work has
  been funded by the Deutsche Forschungsgemeinschaft (DFG, German
  Research Foundation) -- Project-ID 279384907 -- SFB 1245.
\end{acknowledgements}

\bibliography{references}
\end{document}